\documentclass[conference]{IEEEtran} 
\IEEEoverridecommandlockouts
\usepackage{cite}
\usepackage{adjustbox}
\usepackage[ruled,vlined]{algorithm2e}
\usepackage{amsmath,amssymb,amsfonts}
\usepackage{algorithmic}
\usepackage{graphicx}
\usepackage{textcomp}
\usepackage[colorlinks=false, pdfborder={0 0 0}, breaklinks]{hyperref}
\usepackage{xcolor}
\usepackage{balance}
\usepackage{enumitem}
\usepackage{setspace}
\usepackage{wrapfig}
\usepackage{listings}
\usepackage{color}
\usepackage{caption}
\usepackage{subcaption}

\def\BibTeX{{\rm B\kern-.05em{\sc i\kern-.025em b}\kern-.08em
    T\kern-.1667em\lower.7ex\hbox{E}\kern-.125emX}}

\usepackage{tikz}
\usetikzlibrary{shapes.geometric, arrows}
\usetikzlibrary{fit}
\usetikzlibrary{calc}
\usetikzlibrary{shapes.geometric, arrows}
\tikzstyle{arrow} = [thick,->,>=stealth]

\usepackage{ctable}

\newcommand{\ie}{\emph{i.e.}, }
\newcommand{\eg}{\emph{e.g.}, }

\newcommand{\cf}{\emph{cf.}\xspace}
\newcommand{\HAS}{\emph{HTTP Adaptive Streaming }}

\newcommand{\DCT}{\emph{Discrete Cosine Transform }}

\newcommand{\HLS}{\emph{HTTP Live Streaming }}

\newcommand{\QoE}{\emph{Quality of Experience}}

\newcommand{\caps}{\texttt{CAPS}\xspace}

\begin{document}

\title{Content-adaptive Encoder Preset Prediction for Adaptive Live Streaming}

\author{\IEEEauthorblockN{Vignesh V Menon\IEEEauthorrefmark{1}, Hadi Amirpour\IEEEauthorrefmark{1}, Prajit T Rajendran\IEEEauthorrefmark{2}, Mohammad Ghanbari\IEEEauthorrefmark{1}\IEEEauthorrefmark{3},
 Christian Timmerer\IEEEauthorrefmark{1}
}
\IEEEauthorblockA{\IEEEauthorrefmark{1}
Christian Doppler Laboratory ATHENA, Alpen-Adria-Universit{\"a}t, Klagenfurt, Austria}

\IEEEauthorblockA{\IEEEauthorrefmark{2} Universite Paris-Saclay, CEA, List, F-91120, Palaiseau, France}  
\IEEEauthorblockA{\IEEEauthorrefmark{3}
School of Computer Science and Electronic Engineering, University of Essex, Colchester, UK
}
}

\maketitle

\begin{abstract}
In live streaming applications, a fixed set of bitrate-resolution pairs (known as \textit{bitrate ladder}) is generally used to avoid additional pre-processing run-time to analyze the complexity of every video content and determine the optimized bitrate ladder. Furthermore, live encoders use the fastest available preset for encoding to ensure the minimum possible latency in streaming. For live encoders, it is expected that the encoding speed is equal to the video framerate. An optimized encoding preset may result in \textit{(i)} increased \QoE~(QoE) and \textit{(ii)} improved CPU utilization while encoding. In this light, this paper introduces a \textbf{C}ontent-\textbf{A}daptive encoder \textbf{P}reset prediction \textbf{S}cheme (\caps) for adaptive live video streaming applications. In this scheme, the encoder preset is determined using \DCT (DCT)-energy-based low-complexity spatial and temporal features for every video segment, the number of CPU threads allocated for each encoding instance, and the target encoding speed. Experimental results show that \caps yields an overall quality improvement of 0.83 dB PSNR and 3.81 VMAF with the same bitrate, compared to the fastest preset encoding of the HTTP Live Streaming (HLS) bitrate ladder using x265 HEVC open-source encoder. This is achieved by maintaining the desired encoding speed and reducing CPU idle time.
\end{abstract}

\begin{IEEEkeywords}
Live streaming, Encoder preset, QoE, HEVC.
\end{IEEEkeywords}

\section{Introduction}
\textbf{\textit{Motivation:}}
\HAS (HAS) has become the \textit{de-facto} standard in delivering video content for various clients regarding internet speeds and device types. The main idea behind HAS is to divide the video content into segments and encode each segment at various bitrates and resolutions, called \textit{representations}, which are stored in plain HTTP servers. These representations are stored to continuously adapt the video delivery to the network conditions and device capabilities of the client~\cite{Bentaleb2019}. Traditionally, a fixed bitrate ladder, \eg \HLS (HLS) bitrate ladder\footnote{\label{apple_hls_ref}\href{https://developer.apple.com/documentation/http\_live\_streaming/hls\_authoring\_specification\_for\_apple\_devices}{https://developer.apple.com/documentation/http\_live\_streaming/ hls\_authoring\_specification\_for\_apple\_devices}, last access: Sep 30, 2022.}, is used in live streaming. Furthermore, for every representation, maintaining a fixed encoding speed, which is the same as the video framerate, independent of the video content, is a key goal for a live encoder. Although the compression efficiency (in terms of the obtained quality and bitrate) of the output video is an important metric for the encoder, maintaining the encoding speed takes precedence in the live scenario. This is because a reduction in encoding speed may lead to the unacceptable outcome of dropped frames during transmission, eventually decreasing the \QoE~(QoE)~\cite{pradeep_ref}.

\begin{figure}[t]
    \centering
    \includegraphics[width=0.38\textwidth]{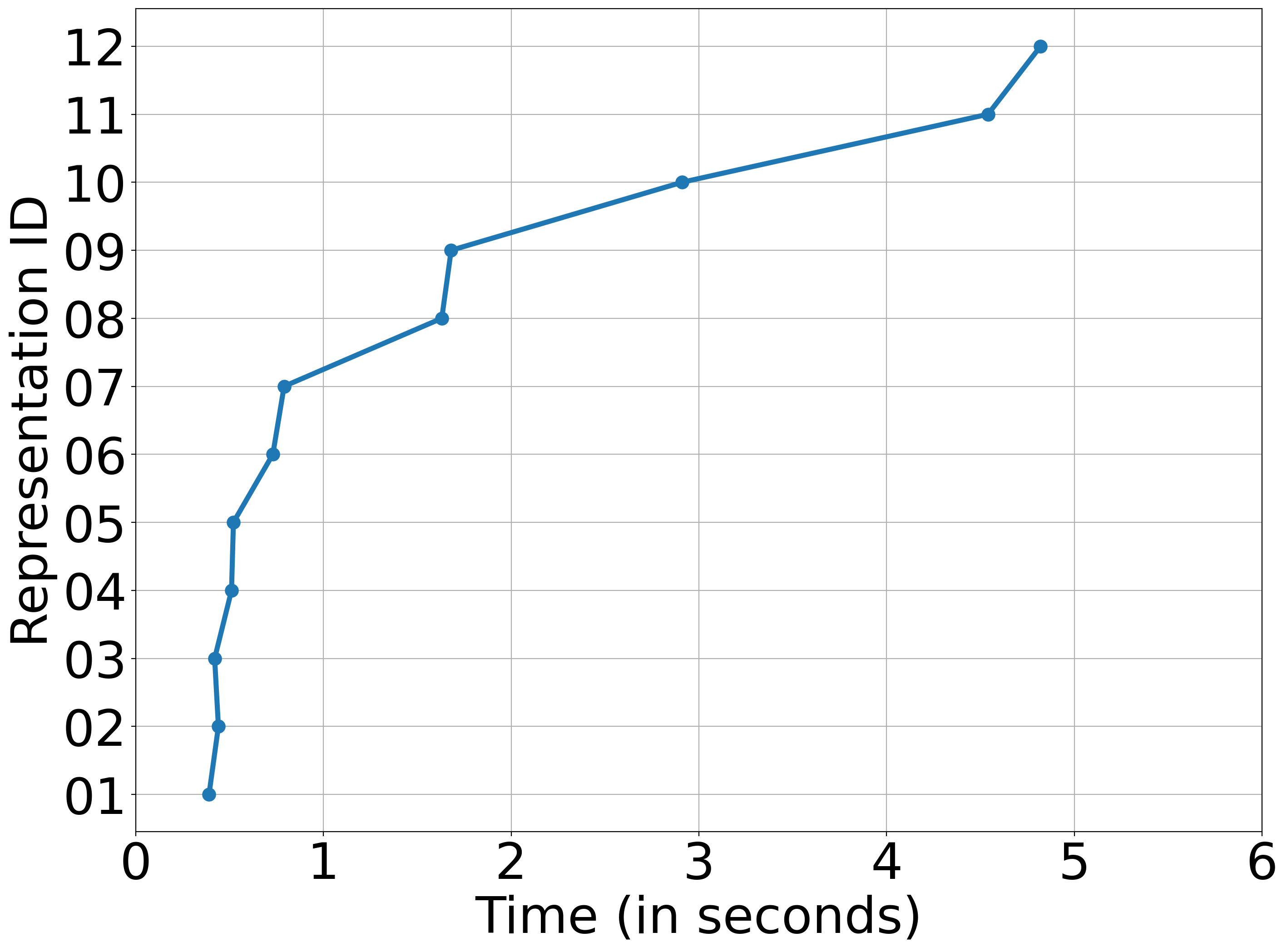}
    \caption{Encoding time of HLS bitrate ladder$^{\ref{apple_hls_ref}}$ representations of the \textit{Wood\_s000} sequence of VCD dataset~\cite{VCD_ref} using \textit{ultrafast} preset of x265$^{\ref{x265_ref}}$.}
    \label{fig:motive_eg}
\end{figure}

\begin{figure*}[t]
\centering
\includegraphics[width=0.80\linewidth]{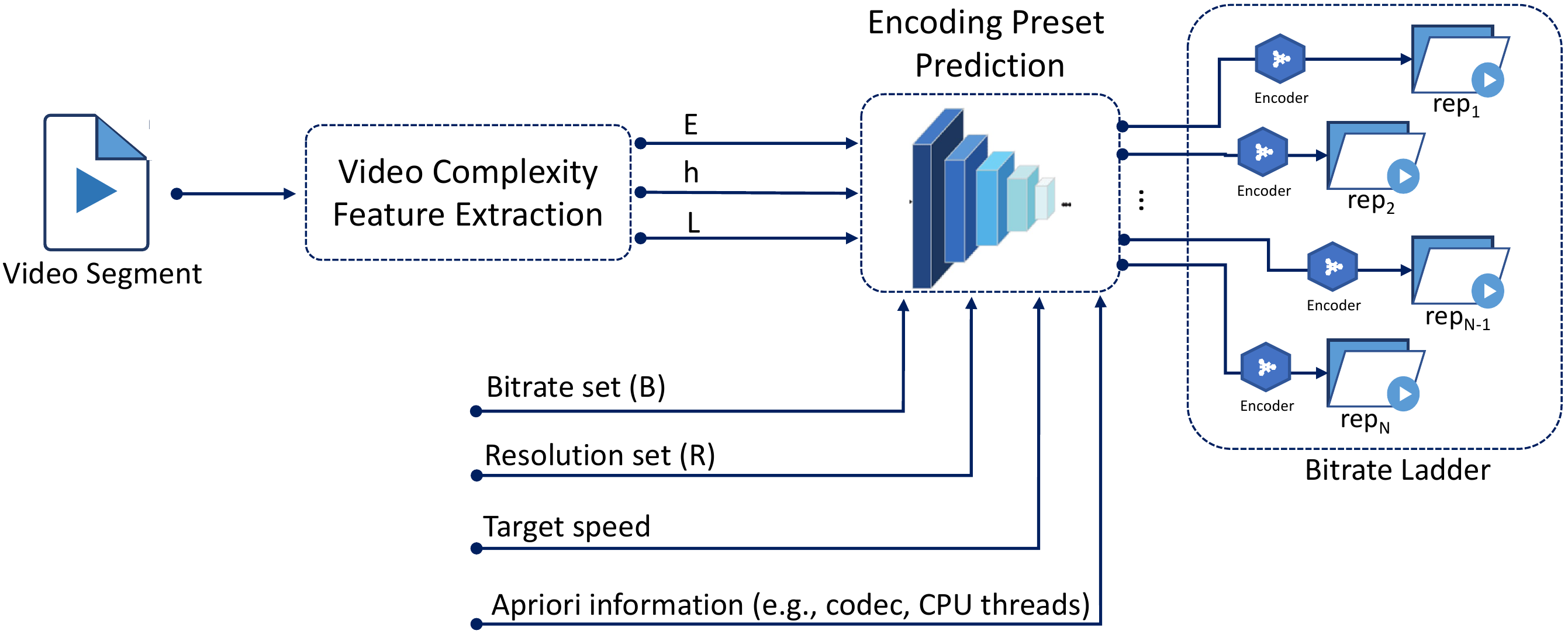}
\caption{The encoding pipeline using \caps envisioned in this paper.}
\label{fig:contribution}
\end{figure*}

Traditional open-source encoders like x264\footnote{\href{https://www.videolan.org/developers/x264.html}{https://www.videolan.org/developers/x264.html}, last access: Sep 30, 2022.}, and x265\footnote{\label{x265_ref}\href{https://www.videolan.org/developers/x265.html}{https://www.videolan.org/developers/x265.html}, last access: Sep 30, 2022.} have pre-defined sets of encoding parameters (termed as \textit{presets}) which present a trade-off between the encoding time and compression efficiency\footnote{\href{https://ottverse.com/choosing-an-x265-preset-an-roi-analysis/}{https://ottverse.com/choosing-an-x265-preset-an-roi-analysis/}, last access: Sep 30, 2022.} . The preset for the fastest encoding (\textit{ultrafast} for x264 and x265) is then used as the encoding preset for all live content, independent of the dynamic complexity of the content. Though this conservative technique achieves the intended result of achieving a live encoding, the resulting encode is sub-optimal, especially when the type of the content is dynamically changing, which is the typical use-case for live streams~\cite{perf_ref}. Furthermore, when the content becomes easier to encode (\ie slow-moving videos or videos that have simpler textures are easy to encode as predicting the current frame from a previous frame is simpler, resulting in smaller residuals), the encoder would achieve a higher encoding speed than the target encoding speed. This, in turn, introduces unnecessary CPU idle time as it waits for the video feed. If the encoder preset is configured such that this higher encoding speed can be reduced while still being compatible with the expected live encoding speed, the quality of the encoded content achieved by the encoder can be improved. Subsequently, when the content becomes complex again, the encoder preset need to be reconfigured to move back to the faster configuration that achieves live encoding speed~\cite{sota_ref1}. 

Furthermore, the encoding speed also depends on the encoding resolution and bitrate. Fig.~\ref{fig:motive_eg} shows the encoding time measurement of HLS bitrate ladder encoding of the \textit{Wood\_s000} sequence~\cite{VCD_ref} using \textit{ultrafast} preset of x265$^{\ref{x265_ref}}$ and eight CPU threads. Since the video segment has a duration of 5 seconds, the target encoding time ($T$) is 5 seconds~\cite{pradeep_ref}. The CPU utilization is poor in the lower bitrate representation encodings; \eg for $0.145$ Mbps to $3.4$ Mbps  representations, an encoding time of less than one second is observed. In the adaptive live streaming scenarios, lower bitrates typically select lower resolutions which might need slower presets to achieve the desired target encoding speed while utilizing the CPU more efficiently.

\textbf{\textit{Goal:}} This paper aims to present an encoding scheme that determines the encoding preset configuration dynamically, adaptive to the video content, to maximize the CPU utilization and maximize the efficiency of adaptive live streaming for a given target encoding speed.

\textbf{\textit{Contributions:}} In this paper, a content-adaptive encoder preset prediction scheme (\caps) is proposed for live video streaming applications to provide low latency streaming while reducing the CPU idle time. To this light, content-aware features, \ie \textit{(i)} \DCT (DCT)-energy-based low-complexity spatial and temporal features, are extracted to determine video segments' characteristics. \textit{(ii)} Based on these features, encoding time is predicted for every preset supported by the encoder using XGBoost~\cite{chen_xgboost_2016} models. Using this information, \emph{optimized encoder presets are determined to maintain the target encoding speed}.

\textbf{\textit{Paper outline:}} In Section~\ref{sec:caps_framework}, the proposed \caps scheme is explained. In Section~\ref{sec:evaluation}, the scheme's performance is validated, and the corresponding experimental results are presented. Finally, Section~\ref{sec:conclusion_future_dir} concludes the paper.

\section{\caps architecture}
\label{sec:caps_framework}
The architecture of \caps for streaming applications is presented in Fig.~\ref{fig:contribution}, according to which the encoder preset for each video segment is predicted using the spatial and temporal features (\ie $E$, $h$, and $L$) of the video segment, the target video encoding speed ($f$), the number of CPU threads used for encoding ($c$) and the set of pre-defined resolutions ($R$) and bitrates ($B$) of the bitrate ladder. The encoding process is carried out with the predicted optimized encoder preset from the set of supported presets ($P$) for each video segment. \caps is classified into two steps: \textit{(i)} video complexity feature extraction and \textit{(ii)} encoder preset prediction, explained in Section~\ref{sec:features}, and Section~\ref{sec:preset_pred}, respectively.

\begin{figure*}[t]
\centering
    \includegraphics[width=0.66\textwidth]{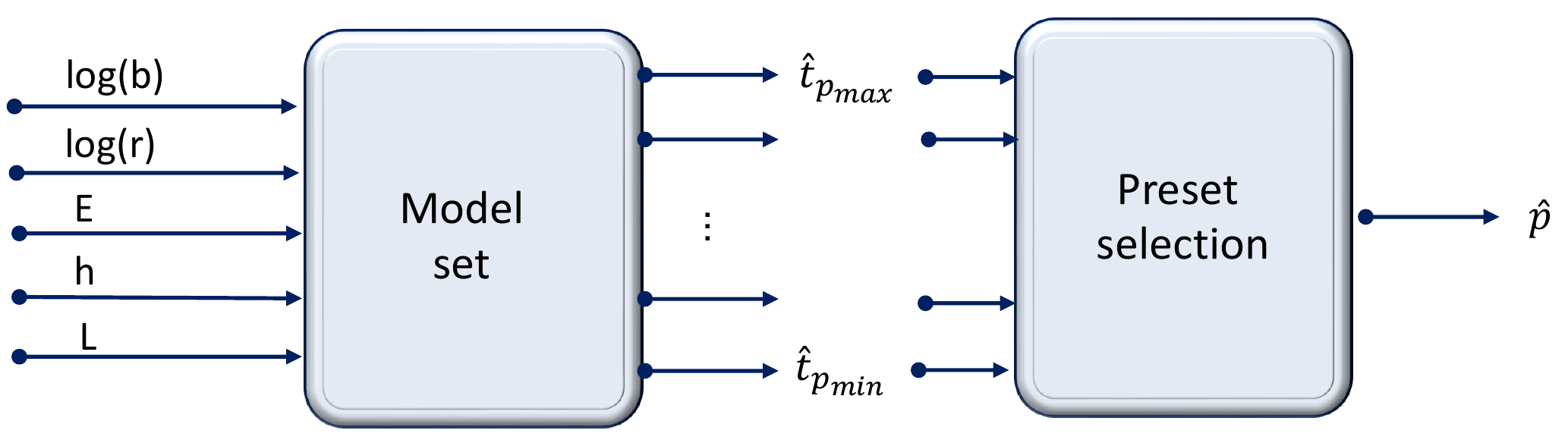}
    \caption{Encoding Preset Prediction for a video framerate ($f$) and the number of CPU threads ($c$).}
\label{fig:cnn_struct}
\end{figure*}

\subsection{Video Complexity Feature Extraction}
\label{sec:features}
In live streaming applications, selecting low-complexity features is critical to ensure low-latency video streaming without disruptions. In this paper, the optimized encoder preset ($\hat{p}$) is determined as a function of three DCT-energy-based features~\cite{dct_ref}, \textit{(i)} the average texture energy ($E$), \textit{(ii)} the average gradient of the texture energy ($h$), and \textit{(iii)} the average luminescence ($L$)~\cite{icip_paper_ref, mmsp_paper_ref,opte_ref, coda_ref, ppte_ref}, the target bitrate ($b$), resolution ($r$), the target video encoding speed ($f$) and the number of CPU threads used for encoding ($c$).

The three DCT-energy-based features are determined as follows. The block-wise texture of each frame is defined as:
\begin{equation}
H_{s, k} = \sum_{i=0}^{w-1} \sum_{j=0}^{w-1} e^{|(\frac{ij}{w^{2}})^2 -1|}|DCT(i,j)|
\end{equation}
where $k$ is the block address in the $s^{th}$ frame, $w \times w$ pixels is the size of the block, and $DCT (i, j)$ is the $(i, j)^{th}$ DCT component when $i + j > 0$, and 0 otherwise~\cite{texture_metric_ref}.
\begin{equation}
\label{eq:spatial_energy}
E = \sum_{s=0}^{S-1} \sum_{k=0}^{K-1} \frac{H_{s, k}}{S \cdot K \cdot w^{2}}
\end{equation}
where $K$ represents the number of blocks per frame, and $S$ denotes the number of frames in the segment. Furthermore, the average \textit{temporal energy} ($h$) is defined as follows:
\begin{equation}
h = \sum_{s=1}^{S-1} \sum_{k=0}^{K-1} \frac{\mid H_{s, k} - H_{s-1, k}\mid}{(S-1) \cdot K \cdot w^{2}}
\end{equation}
The luminescence of non-overlapping blocks $k$ of $s^{th}$ frame is defined as:
\begin{equation}
\label{eq:block_luma}
L_{s, k} = \sqrt{DCT(0,0)}
\end{equation}
where $DCT(0,0)$ is the $DC$ component in the DCT calculation. The block-wise luminescence is averaged per segment denoted as $L$ as shown below.
\begin{equation}
\label{eq:luma}
L = \sum_{s=0}^{S-1}\sum_{k=0}^{K-1} \frac{L_{p, k}}{P \cdot K \cdot w^{2}}
\end{equation}

\subsection{Encoding Preset Prediction}
\label{sec:preset_pred}
The objective of selecting the optimized encoding preset based on resolution, bitrate, and video complexity is decomposed into two parts: \textit{(i)} the former deals with designing models to predict the encoding time for each corresponding input, and \textit{(ii)} the latter involves developing a function to obtain the optimized preset based on the predicted encoding times for each available encoding preset. As explained above, Fig.~\ref{fig:cnn_struct} illustrates the architecture proposed in this paper to facilitate the encoding preset prediction in a two-step manner. The input to the encoding time prediction models designed for a given encoder, number of CPU threads for encoding ($c$), and target encoding speed ($f$) comprises complexity metrics ($E$, $h$, and $L$) along with the resolution ($r$) and bitrate ($b$). $r$ and $b$ are input in logarithmic scale to the encoding time prediction models to reduce internal covariate shift~\cite{batch_normalize_ref}. The model set is trained to predict the encoding times for the pre-defined set of encoding presets ($P$). The minimum and maximum encoder preset ($p_{min}$ and $p_{max}$, respectively) are chosen based on the target encoder. For example, x265 HEVC~\cite{HEVC} encoder supports encoding presets ranging from 0 to 9 (\ie \textit{ultrafast} to \textit{placebo}). The model set predicts the encoding times for each of the presets in $P$ as $\hat{t}_{p_{min}}$ to $\hat{t}_{p_{max}}$.
The predicted encoding times are fed into the preset selection function, which chooses the optimized preset to ensure that the encoding time matches the target video encoding speed. This is accomplished by selecting the preset which is closest to the target encoding time $T$, which is defined as:
\begin{equation}
    T = \frac{n}{f}
\end{equation}
where n represents the number of frames in the segment. Apart from this, the function should ensure that the encoding time is not greater than the target encoding time $T$. This can be represented mathematically as follows:
\begin{equation}
    \hat{t} = arg min_{p} \mid T - t_{p} \mid \hspace{0.3cm}\textit{c.t.}\hspace{0.2cm} p\in [p_{min}, p_{max}]; \hat{t} \leq T
\end{equation}

where $p$ is the selected optimum preset and $\hat{t}$ is the encoding time for $p$. 

%

\textit{Implementation of prediction models:} In this paper, the XGBoost~\cite{chen_xgboost_2016} algorithm is used to train encoding time prediction models for various resolution and preset combinations. XGBoost models are an implementation of gradient boosted decision trees that boast of a significant level of speed and performance as opposed to other machine learning models. $(p_{max} - p_{min} + 1)\cdot n_{r}$ models are trained, where $n_{r}$ denotes the number of resolutions supported by the streaming service provider. Training prediction models for each resolution and preset combination ensures scalability, as more resolutions and presets can be added to \caps architecture in the future with minimal re-training. The input vector passed to the model set (\cf Fig.~\ref{fig:cnn_struct}) is $[E, h, L, log(r), log(b)]$. Inside the model set, prediction models corresponding to the resolution $r$ is used to predict the encoding times for all presets in $P$ for the target bitrate $b$. Thus, the output from the model set is the predicted encoding time for all presets, \ie $\hat{t_p}_{i}$, $\forall i \in [p_{min}, p_{max}]$. Since the encoding time prediction is a regression problem, the loss function employed is mean absolute error.

\begin{table*}[t]
\caption{Representations considered in this paper$^{\ref{apple_hls_ref}}$.}
\centering
\resizebox{0.86\linewidth}{!}{
\begin{tabular}{l|c|c|c|c|c|c|c|c|c|c|c|c}
\specialrule{.12em}{.05em}{.05em}
Representation ID & 01 & 02 & 03 & 04	& 05 & 06 & 07 & 08 & 09 & 10 & 11 & 12 \\
\specialrule{.12em}{.05em}{.05em}
$r$ (width in pixels) & 360 & 432 & 540 & 540 & 540 & 720 & 720 & 1080 & 1080 & 1440 & 2160 & 2160 \\
$b$ (in Mbps) & 0.145 & 0.300 & 0.600 & 0.900 & 1.600 & 2.400 & 3.400 & 4.500 & 5.800 & 8.100 & 11.600 & 16.800 \\
\specialrule{.12em}{.05em}{.05em} 
\end{tabular}}
\label{tab:hls_ladder}
\end{table*}
\begin{table}[t]
\caption{Average encoding time prediction accuracy results of every preset considered for experimental validation.}
\centering
\resizebox{0.995\linewidth}{!}{
\begin{tabular}{l|c|c|c|c|c|c|c|c|c}
\specialrule{.12em}{.05em}{.05em}
Preset &  0 & 1 & 2 & 3 & 4 & 5 & 6 & 7 & 8 \\
\specialrule{.12em}{.05em}{.05em}
$R^{2}$ & 0.97 & 0.96 & 0.97 & 0.97 & 0.97 & 0.96 & 0.97 & 0.98 & 0.98 \\
MAE     & 0.06 & 0.09 & 0.12 & 0.17 & 0.22 & 0.31 & 0.33 & 0.41 & 0.49 \\
\specialrule{.12em}{.05em}{.05em} 
\end{tabular}}
\label{tab:time_pred_res}
\end{table}

\section{Evaluation}
\label{sec:evaluation}

\subsection{Test Methodology}
\label{sec:test_methodology}
In this paper, four hundred sequences (80\% of the sequences) from the Video Complexity Dataset~\cite{VCD_ref} are used as the training dataset, and the remaining (20\%) is used as the test dataset. The sequences are encoded at 24fps using x265 v3.5$^{\ref{x265_ref}}$, \ie $T$ = 5 seconds. The presets defined in x265: 0 (\textit{ultrafast}) to 8 (\textit{veryslow}) are used for evaluation. All experiments are run on a dual-processor server with Intel Xeon Gold 5218R (80 cores, frequency at 2.10 GHz), where each encoding instance uses 8 CPU threads  (\ie $c$ = 8) with multi-threading and x86 SIMD~\cite{x86_simd_ref} optimizations. The DCT-energy-based features, $E$, $h$, and $L$ are extracted using VCA~\cite{vca_ref} running as a pre-processor using 8 CPU threads with multi-threading and x86 SIMD optimizations. The bitrate ladder, as shown in Table~\ref{tab:hls_ladder} is considered in the evaluation. 

The resulting overall quality in Peak Signal to Noise Ratio (PSNR) and Video Multimethod Assessment Fusion (VMAF)\footnote{\href{https://netflixtechblog.com/vmaf-the-journey-continues-44b51ee9ed12}{https://netflixtechblog.com/vmaf-the-journey-continues-44b51ee9ed12}, last access: Sep 30, 2022.}, and the achieved bitrate are compared for each test sequence. Bjøntegaard Delta values~\cite{DCC_BJDelta} BD-PSNR and BD-VMAF refer to the average increase in PSNR and VMAF of the representations compared with the \textit{ultrafast} preset encoding with the same bitrate, respectively. A positive BD-PSNR and BD-VMAF indicate a gain in coding efficiency of \caps compared to the \textit{ultrafast} preset encoding.


\begin{figure}[t]
    \centering
    \includegraphics[width=0.38\textwidth]{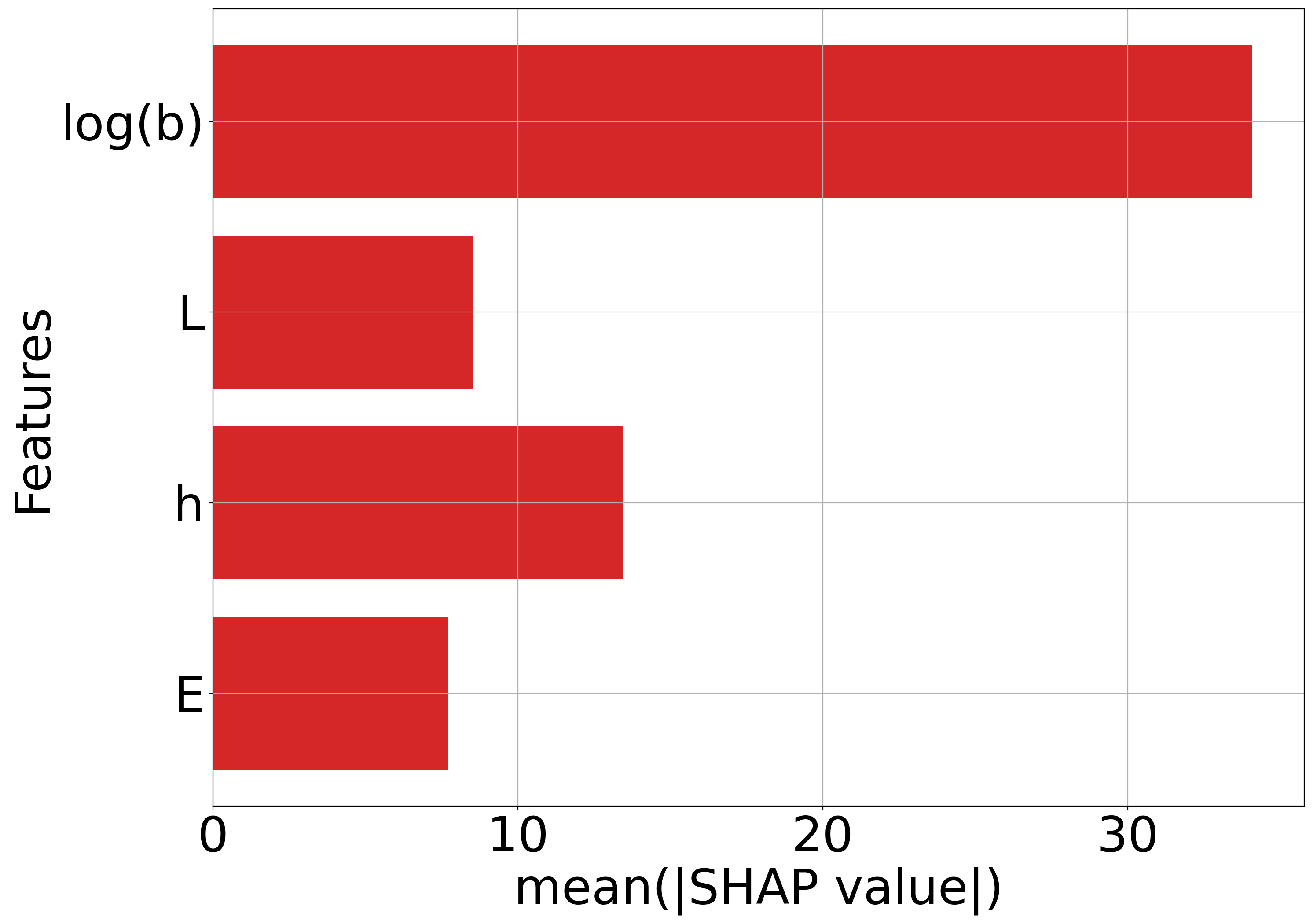}
    \caption{Average relative importance of features in the encoding time prediction of 2160p encoding for all presets.}
    \label{fig:shap_preset}
\end{figure}

\subsection{Experimental Results}
\label{sec:exp_results}

Firstly, the overall time taken to predict optimized preset is evaluated. $E$, $h$, and $L$ features are extracted at an average speed of 370 frames per second. The inference time of the XGBoost model is 0.5ms. Hence, the overall speed of preset prediction is 240 frames per second. Secondly, the encoding time prediction accuracy of the XGBoost models for each preset is observed in terms of $R^{2}$ score and Mean Absolute Error (MAE), as shown in Table~\ref{tab:time_pred_res}. The average $R^{2}$ score and MAE are 0.97 and 0.21, respectively. This paper also evaluates the relative importance of features in the prediction models for each resolution in terms of SHAP values~\cite{shap_ref}, as shown in Fig.~\ref{fig:shap_preset}. The target bitrate in the logarithmic scale ($log(b)$) is the most critical feature for encoding time prediction, followed by the $h$, $L$, and $E$ features. 

\begin{figure}[t]
    \centering
    \includegraphics[width=0.36\textwidth]{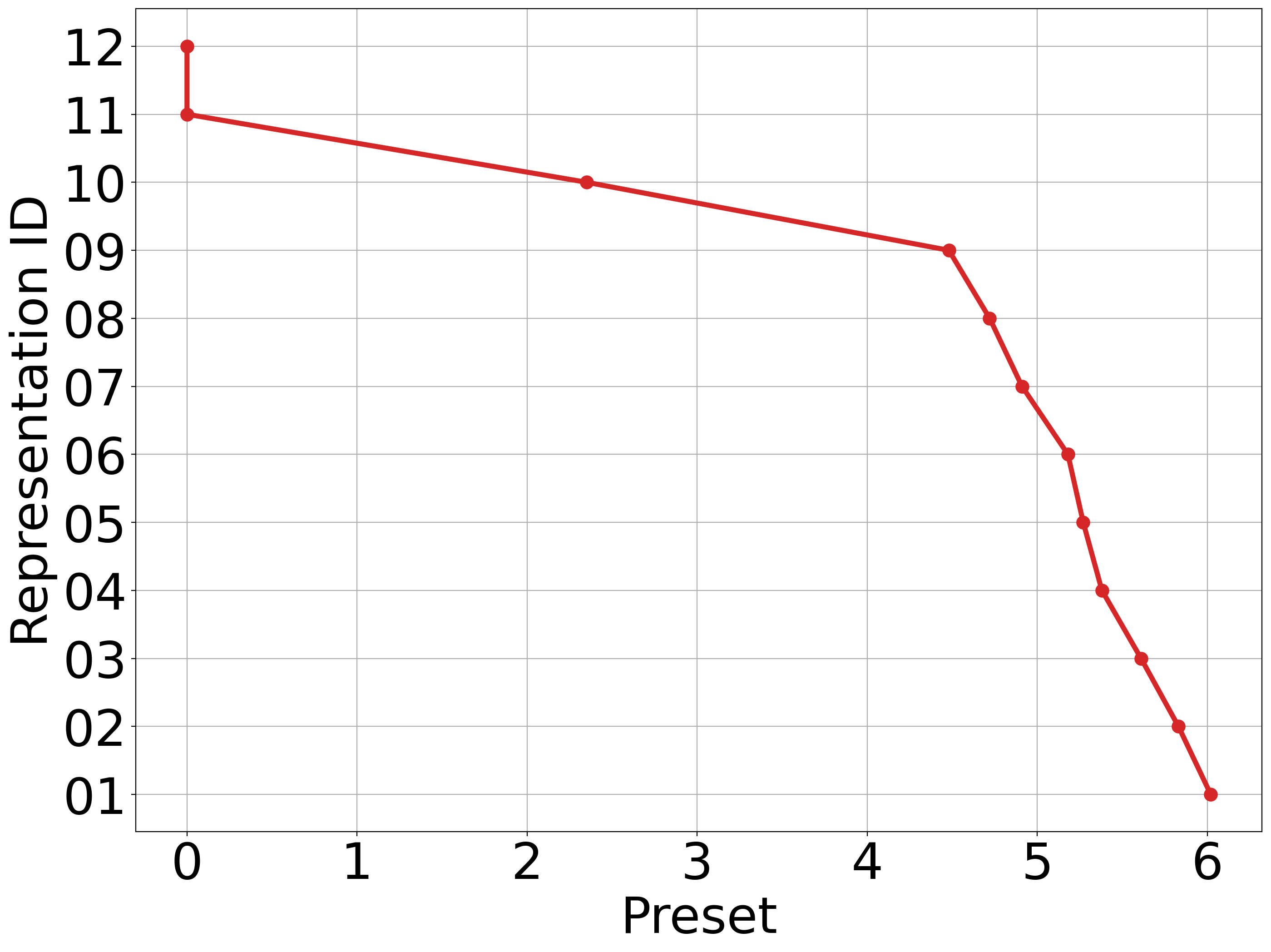}
    \caption{Average preset chosen for each representation in the HLS bitrate ladder.}
    \label{fig:avg_preset}
\end{figure}

\begin{figure*}[t]
\centering
\begin{subfigure}{0.32\textwidth}
    \centering
    \includegraphics[width=\textwidth]{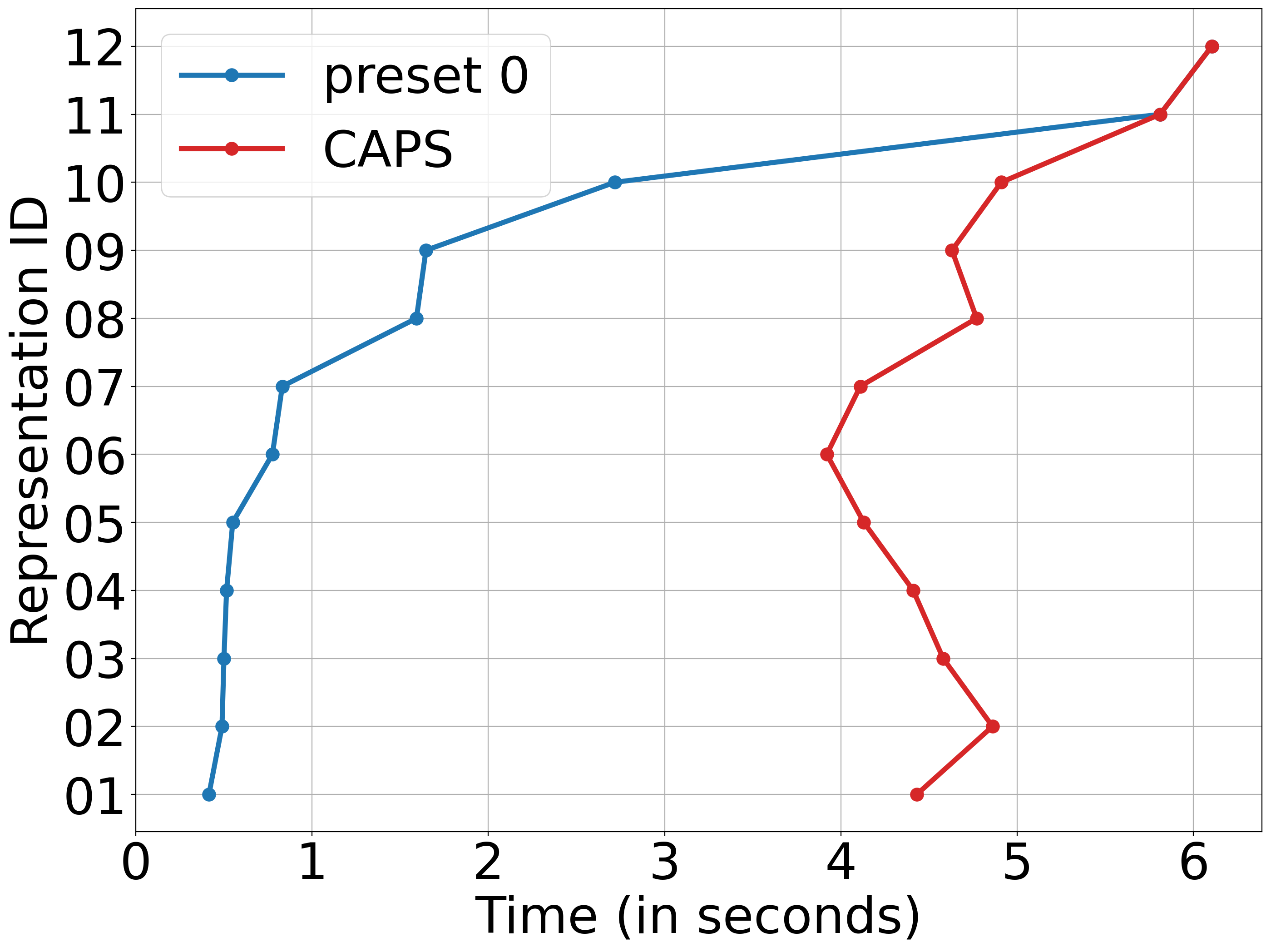}
    \caption{}
        \label{fig:time_rep}
\end{subfigure}
\hfill
\begin{subfigure}{0.32\textwidth}
    \centering
    \includegraphics[width=\textwidth]{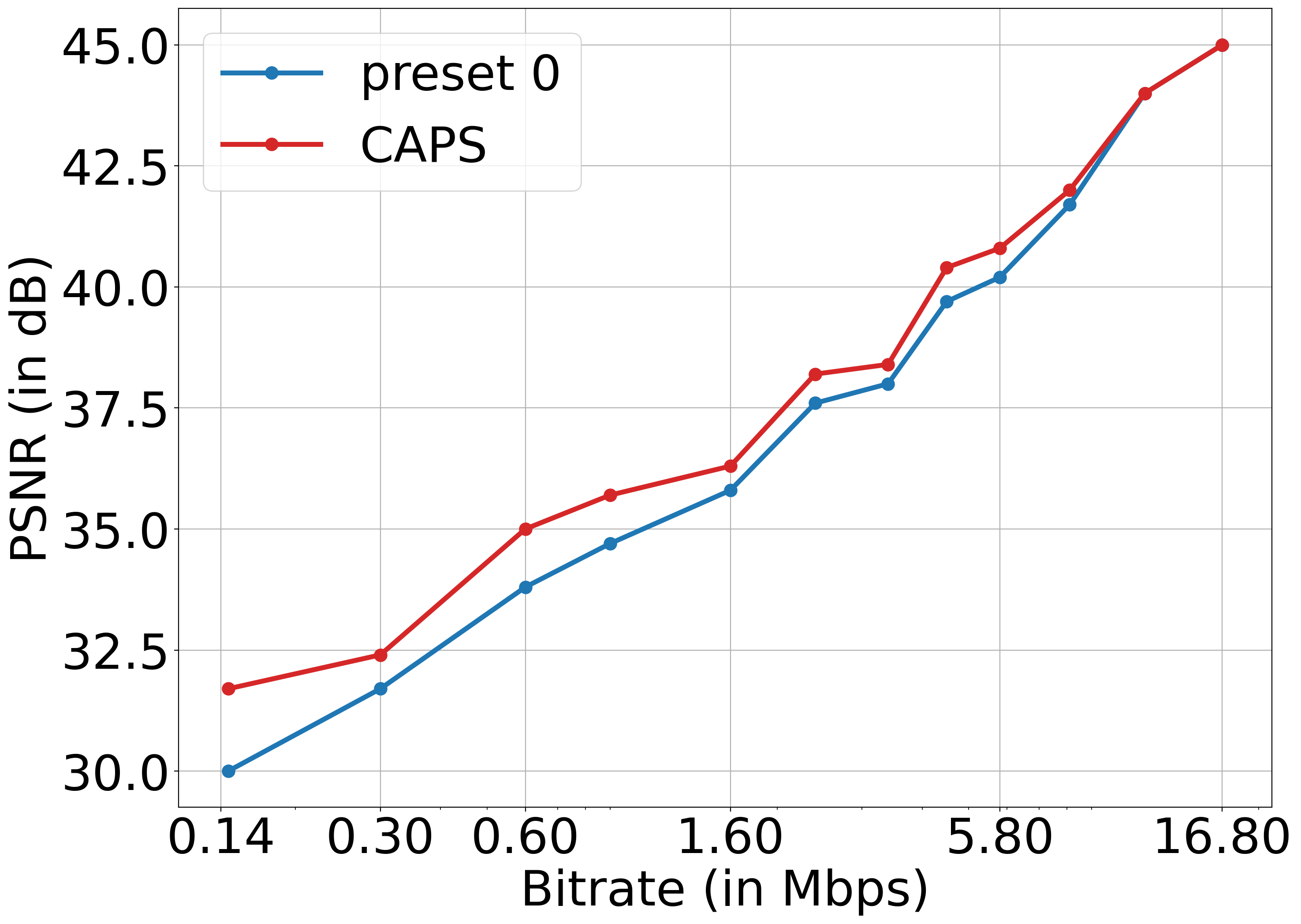}
    \caption{}
        \label{fig:psnr_rep}    
\end{subfigure}
\hfill
\begin{subfigure}{0.32\textwidth}
    \centering
    \includegraphics[width=\textwidth]{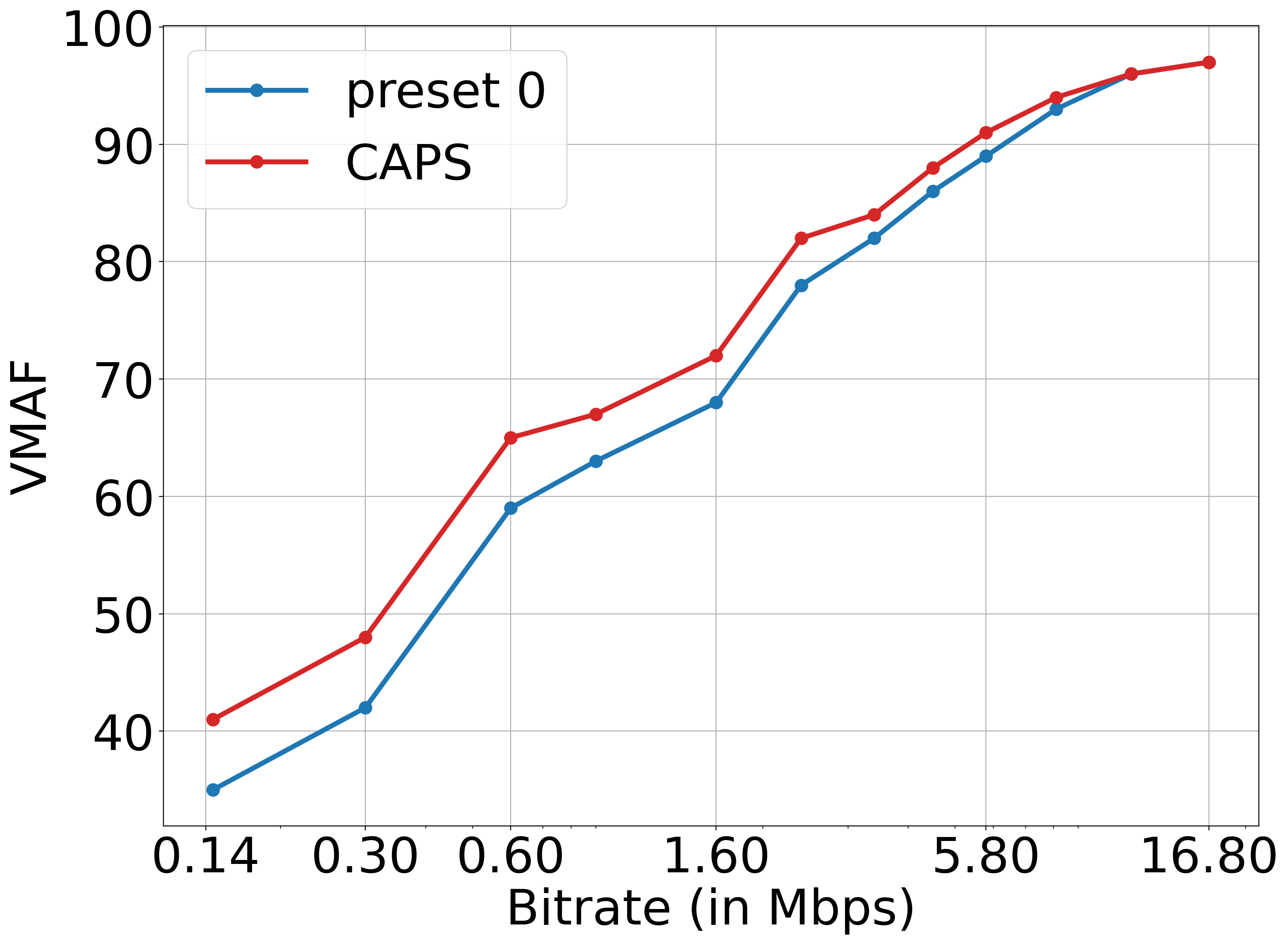}
    \caption{}
    \label{fig:vmaf_rep}    
\end{subfigure}
\caption{\textit{(a)} Average encoding time, \textit{(b)} Average PSNR, and \textit{(c)} Average VMAF for each representation.}
\label{fig:main_plot}
\end{figure*}

Fig.~\ref{fig:avg_preset} shows the average preset chosen for each representation of the HLS bitrate ladder across the test dataset. On average, representation 01 ($0.145$ Mbps) chooses \textit{slow} preset ($p = 6$), while representations 11 ($11.6$ Mbps) and 12 ($16.8$ Mbps) choose \textit{ultrafast} preset ($p = 0$). Fig.~\ref{fig:time_rep} shows the average video encoding time for every representation in the bitrate ladder using \textit{ultrafast} preset and \caps. On average, the $11.6$ Mbps and $16.8$ Mbps representations take more than 5 seconds for encoding using \textit{ultrafast} preset. Hence, these representations cannot be encoded faster than the existing presets. For the other representations, it is observed that the encodings are finished within the bound of $T = 5$. It is also observed that using \textit{ultrafast} preset for all representations introduces significant CPU idle time for lower bitrate representations. However, \caps yield lower CPU idle time when the encodings are carried out concurrently. Fig.~\ref{fig:psnr_rep} and Fig.~\ref{fig:vmaf_rep} show the average PSNR and VMAF for each representation, respectively. It is observed that the visual quality improves significantly at lower bitrate representations. Using \caps yields an overall BD-VMAF of 3.81 and BD-PSNR of 0.83 dB.

\section{Conclusions}
\label{sec:conclusion_future_dir}
This paper proposed \caps, a content-adaptive encoder preset prediction scheme for adaptive live streaming applications. \caps predicts the optimized encoder preset for a given target bitrate, resolution, and video framerate for each segment, which helps improve the quality of video encodings. DCT-energy-based features are used to determine segments' spatial and temporal complexity. The performance of \caps is analyzed using the x265 open-source HEVC encoder for the HLS bitrate ladder encoding. It is observed that \caps yield lower idle time and an overall quality improvement of 0.83dB PSNR and 3.81 VMAF score with the same bitrate, compared to the fastest preset encoding of the reference HTTP Live Streaming (HLS) bitrate ladder. 

\section{Acknowledgment}
The financial support of the Austrian Federal Ministry for Digital and Economic Affairs, the National Foundation for Research, Technology and Development, and the Christian Doppler Research Association is gratefully acknowledged. Christian Doppler Laboratory ATHENA: \url{https://athena.itec.aau.at/}.
\balance
\bibliography{references.bib}{}
\bibliographystyle{IEEEtran}
\balance
\end{document}